\documentclass[prl,twocolumn,showpacs]{revtex4}
\usepackage{graphicx}
\usepackage{color}
\usepackage{amsmath}
\usepackage{tabularx,units}
\usepackage{amssymb}
\usepackage{multirow}
\begin{document}

\title{Quasi-universal transient behavior of a nonequilibrium Mott insulator driven by an electric field.}
\author{K.~Mikelsons and J.~K.~Freericks}
\affiliation{Department of Physics, Georgetown University, Washington, DC, 20057 USA}
\author{H.~R.~Krishnamurthy}
\affiliation{Centre for Condensed Matter Theory, Department of Physics, Indian Institute of Science, Bangalore 560012, India}
\affiliation{Condensed Matter Theory Unit, Jawaharlal Nehru Centre for Advanced Scientific Research, Bangalore 560064, India}

\date{\today}

\begin{abstract}
We use a self-consistent strong-coupling expansion  for the self-energy (perturbation theory in the hopping) 
to describe the nonequilibrium dynamics of strongly correlated lattice fermions. 
We study the three-dimensional homogeneous Fermi-Hubbard model driven by an external electric field 
showing that the damping of the ensuing Bloch oscillations depends on the direction of the field,
and that for a broad range of field strengths, a long-lived transient prethermalized state emerges.
This long-lived transient regime implies that thermal equilibrium may be out of reach 
of the time scales accessible in present cold atom experiments, 
but shows that an interesting new quasi-universal transient state exists in nonequilibrium governed by a thermalized kinetic energy but not a thermalized potential energy.
In addition, when the field strength is equal in magnitude to the interaction between atoms, 
the system undergoes a rapid thermalization,
characterized by a different quasi-universal behavior of the current and spectral function for different values of the hopping.
\end{abstract}

\maketitle

%


\paragraph{Introduction.}


Nearly all experiments with cold atoms involve nonequilibrium 
processes in either the experimental probes used for measurements or in the 
experimental setup itself.
Of particular interest is the nonequilibrium behavior when the interactions 
are strong enough to create a Mott insulator.
In this regime, perturbative methods based on an expansion in the interaction
strength are bound to fail. In this work, we consider 
a different, complementary technique, based on a strong-coupling expansion~\cite{freericks96,freericks09,jordens10,scarola09},
and demonstrate its application to nonequilibrium systems.
The strong-coupling expansion has proved to be an accurate
and well controlled approximation for equilibrium cold atom problems~\cite{freericks09,jordens10} (especially for temperatures larger than the hopping), yielding excellent approximate results at a fraction of 
the computational cost of more exact methods.

In nonequilibrium, it is crucial to account for the damping and relaxation 
effects in a perturbed system.
Typical random phase approximation (RPA) type calculations include 
the lowest-order correction to the self-energy 
and cannot capture these damping effects.  So we must go to the next order and we include all corrections up to the second order in hopping, 
which allows us to capture the damping effects, 
although at an increased computational cost.
Other related techniques for the strong-coupling regime include the
hybridization expansion~\cite{eckstein10b} used in 
dynamical mean-field theory (DMFT), but our approach will work in any dimension.

To illustrate this method, we study Bloch oscillations 
in a Mott insulating cold atom system in an artificial electric field. 
Similar work has been done within the context of DMFT~\cite{freericks06,freericks08,eckstein10a,eckstein11b}.
Here we show that a prethermalized state persists for a long time after 
the transient excitation of the system, and that it shows quasi-universal behavior
for different strengths of the field or hopping
(similar prethermalized states have been studied in other contexts~\cite{moeckel08,moeckel10,eckstein11a}).
We also find a very rapid thermalization when the field equals 
the interaction strength with scaling behavior for the current and for the density of states
for a wide range of hoppings.


\paragraph{Model.}
We describe the nonequilibrium cold atom system
with a single-band Fermi-Hubbard model
for a homogeneous system with no trap potential:
\begin{align}
H(t) =& \sum_i H^{(0)}_i(t) + H^{(hop)}(t) \nonumber \\
H(t) =& \sum_{i} U(t) n_{i\uparrow} n_{i\downarrow} - J(t)\sum_{\langle ij \rangle,\sigma} \left( e^{i\vec{A}(t)\vec{r}_{ij}} c_{i\sigma}^{\dagger}c_{j\sigma}^{\phantom{\dagger}} + h.c.\right) ,
\label{eq:H_Hubbard}
\end{align}
where
$J(t)$ is the hopping amplitude between the nearest neighbor lattice sites,
$U(t)$ is the energy penalty for two atoms to occupy the same optical lattice site,
$\vec{A}(t)$ is the vector potential fully describing the external ``electric" field
 $\vec{E}(t)$: $\vec{A}(t) = -\int_0^t \vec{E}(t') dt'$ 
(which corresponds to ``pulling" the optical lattice through the atomic cloud in a cold atom experiment). 
$c_{i\sigma}^{\dagger}$ is the creation operator for an atom on site $i$ in the hyperfine
(``spin") state $\sigma$ with two available spin states: $\sigma = \uparrow,\downarrow$. 
The occupancy at site $i$ in state $\sigma$ is $n_{i\sigma} = c_{i\sigma}^{\dagger}c_{i\sigma}$ and
$\vec{r}_{ij} = \vec{r}_{i} - \vec{r}_{j}$, where $\vec{r}_{i}$ is the position vector of lattice site $i$.
In this model, the hopping $J(t)$, interaction $U(t)$, and field $\vec{E}(t)$ can be time dependent.
The vector potential is used to describe the electric field because it preserves translation invariance.
Due to gauge invariance, the same field can be described 
by an appropriate scalar potential, which corresponds to a ``lattice tilting" by adding a linear potential in a cold atom experiment.

\begin{figure}[h]
\begin{center}
\includegraphics*[width=2.0in]{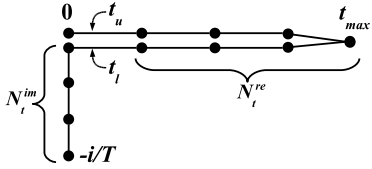}
\caption{The Kadanoff-Baym-Keldysh contour describing the nonequilibrium evolution of the system in real time $t=0 \ldots t_{max}$ 
following an initial (equilibrium) thermalization at temperature $T$. 
Numerical solution requires discretization resulting in $N_t = 2N_t^{re} + N_t^{im}$ finite time steps.}
\label{fig:Keldysh}
\end{center}
\end{figure}

\paragraph{Methodology.}
We solve the model using the strong-coupling expansion~\cite{freericks96,freericks09,scarola09}, 
which treats the hopping amplitude as a small parameter.
The single-particle propagation is fully described
 by the contour-ordered Green's function  
 $G(t,t') = -i \langle T_c c(t) c^{\dag}(t') \rangle$ where both
 time arguments lie on the Kadanoff-Baym-Keldysh contour (Fig.~\ref{fig:Keldysh}).
In the atomic limit, the Green's function on site $i$ is given by 
 $G^{(0)}_i(t,t') = -i \langle T_c c_i(t) c_i^{\dag}(t') \rangle_{H^{(0)}_i}$.
The first-order correction to the full Green's function is due to a single hopping 
to/from a neighboring site:
\begin{equation}
G_{ij}^{(1)}(t,t') = \int_C dt_1 G^{(0)}_i(t,t_1)  J_{ij}(t_1) G^{(0)}_j(t_1,t')\,,
\label{eq:G1}
\end{equation}
where the time integral is taken over the full contour.
All local second-order terms can be expressed as~\cite{suppmatt}:
\begin{align}
\label{eq:G2}
G_{ii}^{(2)}(t,t') =&  \int_C dt_1 \int_C dt_2 \sum_m  J_{im}(t_2) J_{mi}(t_1) \nonumber \\
              & \times  \mathfrak{G}^{(0)}_i (t,t_1,t_2,t')  G^{(0)}_m(t_2,t_1)\,,
\end{align}
where 
\begin{align}
\label{eq:G2_cumulant}
\mathfrak{G}^{(0)}_i (t_0,t_1;t_2,t_3) &= \mathcal{G}^{(0)}_i(t_0,t_1,t_2,t_3) \nonumber \\
                   & + G^{(0)}_i(t_0,t_3)  G^{(0)}_i(t_1,t_2) \nonumber \\
                   & - G^{(0)}_i(t_0,t_2)  G^{(0)}_i(t_1,t_3) 
\end{align}
is a second-order cumulant Green's function and 
$\mathcal{G}^{(0)}_i(t_0,t_1,t_2,t_3)=-i\langle T_c c_i(t_0) c_i(t_1) c_i^{\dag}(t_2) c_i^{\dag}(t_3) \rangle_{H^{(0)}_i}$
is the two-particle atomic Green's function.

\begin{figure}[h]
\begin{center}
\renewcommand{\arraystretch}{1.5}
\begin{tabular}{c c c c c c c c} 
\multirow{2}{*}{a)\hspace{0.1in} } & $G$ & \hspace{0.2in} & $G^{(0)}$ & \hspace{0.2in} & $J_k$ & \hspace{0.2in} & $G^{(2)}_{ii}$ \\        
                    & \includegraphics[scale=0.7]{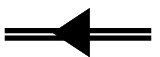} & \hspace{0.2in} & \includegraphics[scale=0.7]{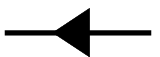} & \hspace{0.2in} & 
\includegraphics[scale=1.0]{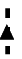} & \hspace{0.2in} & \includegraphics[scale=0.7]{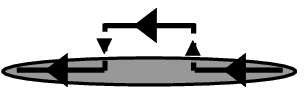} \\ 
b)\hspace{0.1in} &\multicolumn{7}{c}{ \includegraphics*[scale=0.7]{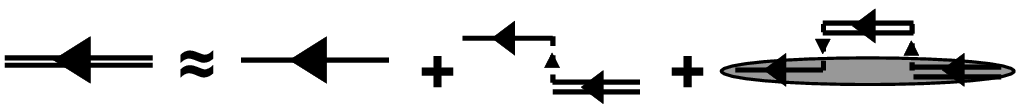} } \\
\end{tabular}
\caption{a) Graphic notation for the full Green's function, 
atomic Green's function, hopping and the second-order local Green's function (Eq.~\ref{eq:G2}).
b) Diagrammatic interpretation of the self-consistency condition. 
The resummation in momentum space requires that the initial propagation in the first and second-order terms
is governed by the full Green's function. 
In addition, we require that the propagation 
on the neighboring site in the second-order term is also the same as the full Green's function. }
\label{fig:Diagrams}
\end{center}
\end{figure}

Due to the increased computational complexity, the terms
beyond second order are truncated. However, a partial but 
infinite resummation of terms is possible. For a homogeneous system, this leads to 
a matrix equation for the momentum-dependent Green's function:
\begin{equation}
\label{eq:Gk}
G_k = \left\{ [G^{(0)}]^{-1} - \delta_c J_k - \Sigma[G] \right\}^{-1} \,,
\end{equation}
where $J_k = -2J(t) \sum_{i=1}^d \cos(k_i)$ 
is the Fourier transform of the hopping on a $d$-dimensional cubic lattice, 
and we have abandoned the time indices in favor of a matrix notation.
This requires discretization of the contour, so that
the matrix size equals the number of time slices on the contour ($N_t$).
We adopt a convention that $G(t,t) = G^>(t,t)$ 
for the equal time Green's function, which results in following
definition of the delta function on the discretized contour
$\delta_{c,ij} = \delta_{i+1,j} - \delta_{i,1}\delta_{j,N_t}$
for $i,j = 1...N_t$. 
Lastly, $\Sigma[G]$ is the ``second-order cumulant self-energy":
\begin{equation}
\label{eq:Sigma}
\Sigma[G] = \left[ G^{(0)}\right]^{-1} G_{ii}^{(2)}  \left[ G^{(0)}\right]^{-1}\,,
\end{equation}
which implicitly depends on the local single-particle Green's function 
on neighboring sites used to calculate $G_{ii}^{(2)}$.
The only non-local second-order term (which corresponds to hopping two sites away) does not appear in 
the expression in Eq.~(\ref{eq:Gk}), since it is recovered by resummation as a product of first-order terms.
Finally, we require that 
the propagation on the adjacent site in the self-energy
functional (Eq.~\ref{eq:Sigma}) is described by the same 
full Green's function as on the original site,
or, equivalently, that the local Green's function obtained through the momentum summation is 
the same as the one used to calculate the self-energy $\Sigma[G]$.
This approximation is 
justified in a homogeneous system and results in a solution
that captures the relaxation effects to a steady state.
It also imposes a self-consistency condition
which requires an iterative solution.
Remarkably, this method also recovers the exact result
 for $U = 0$, since in that case both 
the second-order cumulant Green's function
 and the self-energy vanish. 
A diagrammatic interpretation of the approximation is shown in Fig.~\ref{fig:Diagrams}.

A number of measurable quantities can be obtained directly from the single-particle Green's function.
The momentum distribution is calculated as:  $n_k(t) = i G_k(t_u,t_l)$,
where $t_u$ ($t_l$) denote upper (lower) branches of real time part of the Keldysh contour (Fig.~\ref{fig:Keldysh}).
The kinetic energy and current can be calculated as sums over $k$-space~\cite{turkowski07}:
\begin{align}
e^{kin}(t) & = -\frac{2J(t)}{N_k} \sum_{k}^{N_k} \sum_m^d \cos(k_m - A_m(t)) n_k(t) \label{eq:Ekin} \\
j_{m}(t) & = \frac{2J(t)}{N_k} \sum_k^{N_k}   \sin(k_m - A_m(t)) n_k(t) \,.\label{eq:Jcur}
\end{align}
The potential energy (and double occupancy) 
can be obtained from the equal time derivative of the Green's function
and the previously calculated kinetic energy:
\begin{align}
\left. \frac{\partial G(t,t')}{\partial t} \right|_{t=t'} & =  U(t) \left\langle n_{\uparrow}(t) n_{\downarrow}(t)  -\frac{3}{4}(n_{\uparrow}(t) + n_{\downarrow}(t)) + \frac{1}{2} \right\rangle \nonumber \\
 & + \frac{e^{kin}(t)}{2}\,.
\label{eq:dGdt}
\end{align}
The local retarded Green's function is:
\begin{align}
G_R(t,t') & = -i \Theta(t-t') \langle \left\{ c(t), c^{\dag}(t') \right\} \rangle \nonumber \\
          & = G(t_l,t'_u) - G(t_u,t'_l)\,,
\label{eq:Gret}
\end{align}
and the spectral function at an average time $t_a$ is
 $A_{t_a}(\omega) = -\frac{1}{\pi} \mathrm{Im}\int_0^{\infty} G_R(t_a+t/2,t_a-t/2) e^{i\omega t} dt$.


\paragraph{Results.}
We take the hopping, interaction
and electric field to be time independent. 
We investigate a three dimensional cubic system of $N=32^3$ sites
at half filling (the total number of atoms is equal to the number of optical lattice sites).
The hopping $J/U$ is taken small so that the system is originally in a Mott-insulating state.

\begin{figure}[h]
\begin{center}
\includegraphics*[width=3.2in]{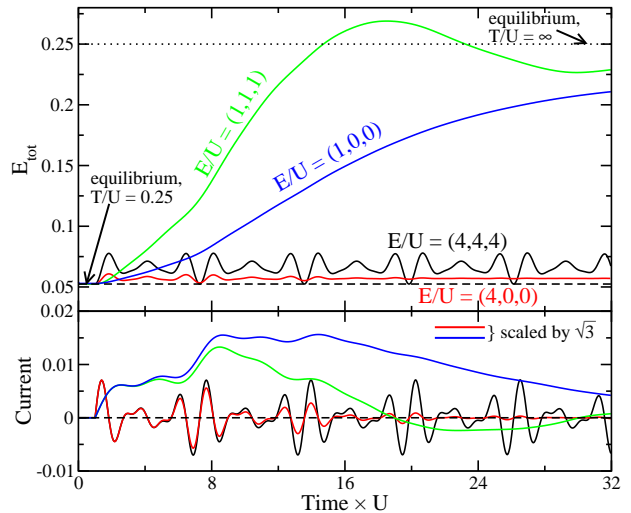}
\caption{(color online) Total energy (top) and current (bottom) for different values of electric field at $U/T = 4$ and $U/J = 24$.
The field is suddenly switched on at $t=1$. 
A strong field results in Bloch oscillations in the current and the total energy, 
yet the damping of the oscillations depends strongly on the direction of the field. 
The dotted line shows the infinite temperature equilibrium limit for the total energy ($E_{tot} = U/4$), 
and the dashed lines correspond to equilibrium with the initial temperature and no field.
The current for cases when the field is aligned with the axial direction of the lattice is scaled by $\sqrt{3}$ for a better
comparison with the results for a diagonal field.}
\label{fig:etot}
\end{center}
\end{figure}

We start with a system initially in thermal equilibrium at a temperature $T/U = 0.25$ 
(corresponding to 6.5\% of sites doubly occupied) and turn on a constant field at time $t=1$.
The resulting response of the energy and the current is shown in Fig.~\ref{fig:etot}.
Sudden application of a strong field ($E>U$) gives rise to Bloch oscillations with a period of $2\pi/E$.
The damping or decay of the Bloch oscillations depends on the number of axial directions for which the field is nonzero.
For a field aligned with the axial direction of the lattice, the Bloch oscillations decay rapidly
with a time-scale determined by the inverse hopping ($J^{-1}$), 
whereas for a field aligned with the diagonal direction, there is essentially no damping within the time scale of the simulation.
A strong field suppresses the tunnelling of particles along the direction of the field,
resulting in an effective two-dimensional system for the axial field, 
and an effective zero-dimensional system for a diagonal field~\cite{kotliar12}.
The amplitude of the current is proportional to the hopping parameter $J$. We show results for one initial temperature here.  As the initial temperature increases, the amplitude of the current decreases. 
The additional amplitude modulation of the current has a dominant frequency of $U$, 
that is independent of field, temperature or hopping.
For both field directions, a strong field results in a small increase of the total energy that is almost entirely 
due to the change in the kinetic energy which is much smaller than the energy in the long-time thermalized state, which will have the infinite temperature limit of the energy. 
Clearly, the time-scale for full thermalization of the system 
far exceeds the time of simulation, as expected, since the creation of double occupancies involves multiple particle processes when the hopping is much smaller than $U$ (due to energy conservation).

\begin{figure}[h]
\begin{center}
\includegraphics*[width=3.2in]{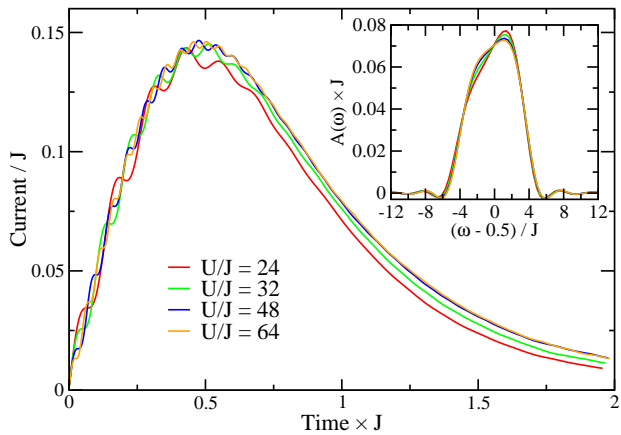}
\caption{(color online) Scaled current for different values of hopping 
for $E = U(1,0,0)$ and $T=J$, showing quasi-universal behavior.
The data change very little with a further decrease of the initial temperature.
The exponential decay of the current has a time-scale of $J^{-1}$. 
The inset shows scaled spectra (upper Hubbard band) at an average time $t_a = 1/J$, 
which also show a quasi-universal behavior.}
\label{fig:current_E1}
\end{center}
\end{figure}

Setting the field equal to the interaction ($E=U$) leads to markedly different behavior 
illustrated by a rapid rise of total energy that approaches (or even overshoots)
the maximum ($T\rightarrow \infty$) value for both directions of the field. 
The current shows a peak that is followed by a slow decay with a time-scale on the order of $J^{-1}$. 
No Bloch oscillations are present, 
but a slight period $2\pi/U$ modulation is apparent during the initial rise of the current. 
Scaling of the current and time for different values of hopping 
reveals the quasi-universal nature of the decay (see Fig.~\ref{fig:current_E1}).
The overall shape of the response depends on the direction of the field
and the dimensionality of the system.
Smaller fields ($E<U$) again lead to Bloch oscillations, 
which become less pronounced with a further decrease of the field~\cite{eckstein10a}.

\begin{figure}[h]
\begin{center}
\includegraphics*[width=3.2in]{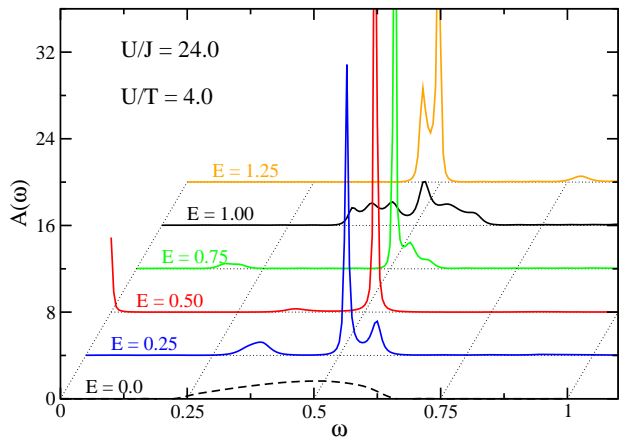}
\caption{(color online) Spectral function at an average time $t_a=32.0$ 
for $U/J=24$ and $U/T=4$ for a diagonal field. 
Due to the even symmetry of the spectral function for the half filled system, only positive energies are shown. 
The application of a field results in a sharp peak near $\omega/U = 0.5$ in all cases, except when $E/U = 1.0$. 
For $E/U=0.5$, a zero frequency peak emerges. See Ref.~\cite{suppmatt} for details on numerics.}
\label{fig:Aw}
\end{center}
\end{figure}

We show the spectral function for the prethermalized state in Fig.~\ref{fig:Aw} for a diagonal field of variable strength.
Even for a small field, the spectral function rapidly collapses into sharp peaks near $\omega/U = \pm 0.5$ 
with smaller sidebands split off the main peaks roughly by plus or minus the size of the electric field.
For $E/U=0.5$, a resonant peak emerges at zero frequency, similar to results
obtained by a numerical renormalization group method in~\cite{joura08}.
In the $E \approx U$ case, the Hubbard bands widen due to rapid thermalization.
Since the retarded Green's function data are restricted to short time intervals, 
we use a method similar to the one introduced by Sandvik~\cite{sandvik98}
to transform data to the frequency domain (see also \cite{suppmatt}). 

When the field is aligned with the lattice, the Bloch oscillations are rapidly damped
and the Hubbard bands remain wide and featureless~\cite{suppmatt}, 
except for $E/U=0.5$, which shows a spectral weight transfer 
from $\omega/U = \pm0.5$ to a zero frequency resonance~\cite{joura08}.


The prethermalized state persists for long times (longer than the inverse hopping $J^{-1}$), 
and unfortunately we cannot directly measure the associated time-scale which is probably exponential in $U/J$~\cite{eckstein10a} or longer,
and has been suggested as related to the slow creation of doublons~\cite{strohmaier10,sensarma10}. 
In a strong field, all our observations are consistent with a dimensional reduction mechanism suggested in~\cite{kotliar12}. 


\paragraph{Conclusions.}
Based on a strong-coupling nonequilibrium method,
we have established the generic presence of a long prethermalized regime in a 
Mott insulator for a wide range of uniform field strengths. 
While damping of Bloch oscillations strongly depends 
on the direction of the field relative to the axial directions of the lattice,
fast thermalization occurs only when field strength approaches the repulsive interaction between atoms.
In this case, the time scale of the thermalization is defined by the hopping, 
and several measurements, including the single-particle spectrum,
point to quasi-universal behavior, which is independent of 
hopping or temperature (within a certain range).
For field strengths different from the interaction, the heating is strongly suppressed 
and the system enters a prethermalized regime characterized by a long lifetime 
which unfortunately cannot be computationally resolved.
The striking difference in the time scales of thermalization 
for different field strengths are experimentally relevant.
The strong-coupling method introduced in this work is quite general and can be used to study a wide range of 
nonequilibrium phenomena. 
Despite an inability to reach the lowest temperatures, 
it is especially well suited to study cold atom systems, 
where current experiments have similar limitations.
The method is also easy to extend for studies of bosonic systems.
In future work, we plan to generalize it to include the trap potential.


\paragraph{Acknowledgments.}
We acknowledge useful discussions with P. Zoller.
This work was supported by a MURI grant from 
Air Force Office of Scientific Research numbered FA9559-09-1-0617. 
Supercomputing resources came from a challenge grant of the DoD at the Arctic Region Supercomputing Center,
the Engineering Research and Development Center and the Air Force Research and Development Center.
The collaboration was supported by the Indo-US Science and Technology Forum under the joint center numbered JC-18-2009 (Ultracold atoms).
JKF also acknowledges the McDevitt bequest at Georgetown and HRK also acknowledges support of the Department of Science and Technology in India.



\begin{thebibliography}{99}


\bibitem{freericks96} J.~K.~Freericks and H.~Monien, Phys.~Rev.~B {\bf 53}, 2691 (1996).

\bibitem{freericks09} J.~K.~Freericks, H.~R.~Krishnamurthy, Yasuyuki Kato, Naoki Kawashima and Nandini Trivedi,
Phys.~Rev.~A {\bf{79}}, 053631 (2009).

\bibitem{jordens10} R.~Jordens {\it et al.} Phys.~Rev.~Lett. {\bf 104}, 180401 (2010).
\bibitem{scarola09} V.~W.~Scarola, L.~Pollet, J.~Oitmaa, and M.~Troyer,  Phys.~Rev.~Lett. {\bf 102}, 135302 (2009).

\bibitem{eckstein10b} M.~Eckstein and Ph.~Werner, Phys.~Rev.~B {\bf{82}}, 115115 (2010).

\bibitem{freericks06} J.~K.~Freericks, V.~M.~Turkowski, and V.~Zlatic, Phys.~Rev.~Lett. {\bf 97}, 266408 (2006).
\bibitem{freericks08} J.~K.~Freericks, Phys.~Rev.~B  {\bf{77}}, 075109 (2008).

\bibitem{eckstein10a} M.~Eckstein, T.~Oka and Ph.~Werner, Phys.~Rev.~Lett. {\bf{105}}, 146404 (2010).

\bibitem{eckstein11b} M.~Eckstein and Ph.~Werner, Phys.~Rev.~Lett. {\bf{107}}, 186406 (2011).

\bibitem{moeckel08} M.~Moeckel and S.~Kehrein, Phys.~Rev.~Lett. {\bf 100}, 175702 (2008).

\bibitem{moeckel10} M.~Moeckel and S.~Kehrein, New~J.~Phys. {\bf 12}, 055016 (2010).

\bibitem{eckstein11a} M.~Eckstein, M.~Kollar, and Ph.~Werner, Phys.~Rev.~Lett. {\bf{103}}, 056403 (2009).

\bibitem{suppmatt} Please see the Supplemental Material for more details.

\bibitem{turkowski07}  V.~M.~Turkowski and J.~K.~Freericks, 
Phys.~Rev.~B {\bf{75}}, 125110 (2007).

\bibitem{kotliar12} C.~Aron, G.~Kotliar and C.~Weber, Phys.~Rev.~Lett. {\bf 108}, 086401 (2012).

\bibitem{sandvik98} A.~W.~Sandvik, Phys.~Rev.~B {\bf{57}}, 10287 (1998).

\bibitem{joura08} A.~V.~Joura, J.~K.~Freericks and Th.~Pruschke,  Phys.~Rev.~Lett. {\bf 101}, 196401 (2008).

\bibitem{strohmaier10} N.~Strohmaier {\it et al}, Phys.~Rev.~Lett. {\bf 104}, 080401 (2010).

\bibitem{sensarma10} R.~Sensarma {\it et al}, Phys.~Rev.~B {\bf 82}, 224302 (2010).









\end{thebibliography}
\end{document}